\documentstyle[prl,floats,twocolumn,aps,epsf,psfrag]{revtex}

\tighten
\begin{document}
\draft
\twocolumn[\hsize\textwidth\columnwidth\hsize\csname @twocolumnfalse\endcsname
\title{Squeezing as an irreducible resource}

\author{Samuel L.~Braunstein}
\address{SEECS, University of Wales, Bangor LL57 1UT, UK}
\address{Hewlett-Packard Labs, Math Group, Bristol BS34 8QZ, UK}

\date{\today}
\maketitle

\begin{abstract}
We show that squeezing is an irreducible resource which remains invariant
under transformations by linear optical elements. In particular, we give
a decomposition of any optical circuit with linear input-output relations
into a linear multiport interferometer followed by a unique set of single
mode squeezers and then another multiport interferometer. Using this
decomposition we derive a no-go theorem for superpositions of macroscopically 
distinct states from single-photon detection. Further, we demonstrate the
equivalence between several schemes for randomly creating 
polarization-entangled states. Finally, we derive minimal quantum optical 
circuits for ideal quantum non-demolition coupling of quadrature-phase 
amplitudes.
\end{abstract}
\vspace{3ex}
]

There is still no consensus as to the eventual working material which
will be used by large scale quantum computers to store and process quantum 
information. By contrast, there seems to be no dispute about using
optical or near-infrared photons for quantum communication. The advantages 
are obvious: high-speed transmission, weak coupling to the environment,
negligible thermal noise. Some disadvantages include the difficulties
in coupling light to light and in creating suitable input states. Some
proposals involve cavity QED \cite{qcomm1,qcomm2,qcomm3}. However, to date 
{\it all\/} implementations of quantum communication protocols (over 
distances larger than microns) have used only coherent state inputs and 
optical components which are no more non-linear than parametric 
down-converters or photodetectors. Thus, outside detection, this suggests 
that near-future quantum communication experiments will also involve 
information processing which can be described by at most a, possibly 
time-dependent, linear mixing of annihilation and creation operators (linear 
Bogoliubov transformations corresponding to quadratic interactions) for 
optical modes. In this paper we will demonstrate that for such systems 
{\it squeezing\/} \cite{qo} forms an irreducible resource which allows us 
to quantify their power.

It has been known for some time how to analytically and numerically 
calculate the evolution of systems under the action of linear Bogoliubov 
transformations \cite{Malkin,Mosh,Ekert}. Here, however, we develop a
tool which will allow us to predict the strengths and limitations of 
devices (and resources). To that end, we would like to formalize our 
equations in terms of a universal set of irreducible resources and a 
restricted set of operations.

As a first step, we will see that any optical system that is modeled by 
linear Bogoliubov transforms can be decomposed into strictly `linear' and 
strictly `non-linear' components. For photonic modes, 
quantum optics provides a well-developed correspondence between laboratory 
components and theoretical mode couplings. In this correspondence, 
traditional optics involves only linear elements (beam-splitters, mirrors,
half-wave plates, etc.). Mathematically, linear optical components have 
Bogoliubov transformations given by
\begin{equation}
\hat b_j = \sum_k U_{jk} \hat a_k \;,
\end{equation}
where $U$ is an arbitrary unitary matrix and there is no mixing of
the mode annihilation and creation operators. Any such unitary $U$
may be explicitly constructed from linear optical primitive
components \cite{Reck}.

By contrast, {\it non-linear optical\/} components (in particular 
squeezers, parametric amplifiers and down-converters) are used to 
generate quantum resources (squeezed states, entangled states, etc). 
These non-linear components may produce a {\it linear\/} mixing between 
annihilation and creation operators when some pumping field or fields are 
strong enough that their quantum fluctuations and pump depletion may be 
neglected (the so-called parametric approximation). It is this regime of 
linear transformations on (photonic) modes that is of interest to us. 
Without attempting to be exhaustive we shall explicitly label three types 
of non-linear optical elements which yield linear Bogoliubov 
transformations:\hfill\break
{\bf Squeezers (S)}: single-mode down-converters (also known 
as parametric amplifiers) may be described by an interaction Hamiltonian of
the form
\begin{equation}
\hat H_{\rm int} = i{r\over 2}(\hat a_1^{\dagger 2} - \hat a_1^2)
\label{paramp} \;,
\end{equation}
here $r$ is the squeezing parameter and we drop extraneous phases from
our descriptions without loss of generality.\hfill\break
{\bf Two-mode down-converters (D${}_2$)}: may be described by
\begin{equation}
\hat H_{\rm int} \propto i(\hat a_1^\dagger \hat a_2^\dagger - 
\hat a_1 \hat a_2) \label{2mode} \;.
\end{equation}
{\bf (Entangling) Four-mode down-converters (E${}_4$)}:
\begin{equation}
\hat H_{\rm int} \propto i(\hat a_1^\dagger \hat a_2^\dagger +
\hat a_3^\dagger \hat a_4^\dagger- \hat a_1 \hat a_2- \hat a_3 \hat a_4) 
\label{4mode} \;.
\end{equation}
These latter devices may be thought of as {\it entangling\/} down-converters 
if, for example, the even (odd) numbered modes represent differing 
polarization states for a mode heading left (right). 

We are now in a position to describe the reduction of linear 
Bogoliubov transformations. This reduction is given by the so-called 
Bloch-Messiah theorem for bosons (a formal extension of the original result 
for fermions in \cite{BM}; see appendix for a compact proof), which states:

\noindent
{\bf Theorem (Bloch-Messiah reduction)}: 
For a general linear unitary Bogoliubov
transformation of the form
\begin{equation}
\hat b_j = \sum_k (A_{jk} \hat a_k +B_{jk} \hat a^\dagger_k) +\beta_j 
\label{BT} \;,
\end{equation}
where $\hat a_j$, $\hat b_j$ are bosonic annihilation operators,
the matrices $A$ and $B$ may be decomposed into a pair of unitary
matrices $U$ and $V$ and a pair of non-negative diagonal matrices
$A_D$ and $B_D$ satisfying
\begin{equation}
A_D^2=B_D^2+{\openone} \label{1moderel} \;,
\end{equation}
with ${\openone}$ the identity matrix, by
\begin{equation}
A= UA_D V^\dagger ~~~~~~~~~~
B= UB_D V^{\rm T} \label{glub} \;.
\end{equation}

\begin{figure}[thb]
\begin{psfrags}
\psfrag{GLUB}[bc]{\,~~~~~~Bloch-Messiah}
\psfrag{reduction}[bc]{~~~~~~~~reduction}
\psfrag{interferometer}[cb]{\small ~~~~~~~~~~~~multi-port}
\psfrag{interferometer U}[bc]{\small ~~~~~~~~~~~~$U$ multi-port}
\psfrag{interferometer V}[bc]{\small ~~~~~~~~~~~~~$V^\dagger$ multi-port}
\psfrag{dc2}[bc]{~~~D${}_2$}
\psfrag{dc4}[bc]{~~\,E${}_4$}
\psfrag{S}[bc]{ ~~S}
\epsfxsize=3.2in
\epsfbox[-10 -10 430 320]{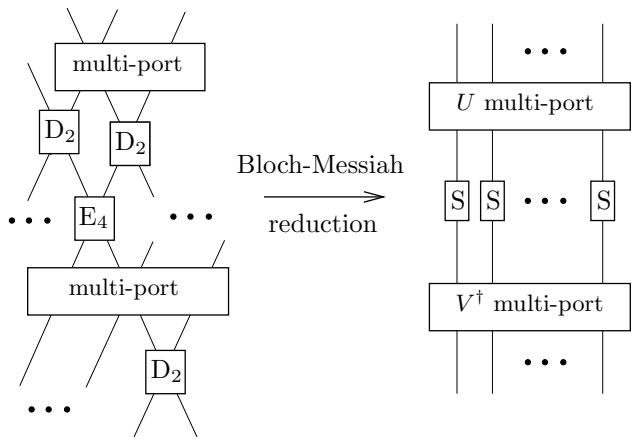}
\end{psfrags}
\caption{An arbitrarily complicated combination of linear multi-port
interferometers, squeezers, down-converters, etc (S, D${}_2$, E${}_4$
etc), each component describable by a quadratic interaction may be 
decomposed by Bloch-Messiah reduction into a linear multi-port described by 
$V^\dagger$, a parallel set of single-mode squeezers (S) and a second 
linear multi-port $U$.}
\label{fig1}
\end{figure}

\noindent
{\bf Corollary}: For optical modes, Bloch-Messiah reduction says that the
general form of multimode evolution with linear Bogoliubov transformations
may be decomposed into a multi-port linear interferometer, followed by 
the parallel application of a set of single-mode squeezers followed by 
yet another multi-port linear interferometer. This reduction is 
illustrated in Fig.~\ref{fig1}.

\noindent
{\bf Remark}: After Bloch-Messiah reduction, the modes acted upon by the
single-mode squeezers will in general involve linear combinations of 
different frequency fields. Thus, the optical circuits so obtained
do not necessarily correspond to an immediate physical decomposition.
Such a physical decomposition may be readily obtained, though it
will no longer have the minimal Bloch-Messiah reduced form.

One common application for down-converters is as sources of interesting 
quantum states. We shall use Bloch-Messiah reduction to tell us something 
about how versatile such devices may be. We shall assume that initial 
coherent states can always be cheaply made available after which we may 
use linear optics and down-converters. The simplest way of operating such 
a source is unconditionally for which we state the following result:

\noindent
{\bf Corollary}: Given a set of initially coherent or vacuum states,
an arbitrarily complicated combination of linear multi-port
interferometers, down-converters, squeezers, etc, will
{\bf deterministically generate} {\it only\/} Gaussian states.

There are at least two other modes of state generation which might
be considered of interest:\hfill\break
{\bf Conditional state generation}: where the required state leaves
some part of the apparatus whenever a suitable sequence of photodetection 
events is found in another part. For example, a weakly coupled two-mode 
down-converter~(\ref{2mode}) can make a single-photon 
state to a good approximation in either of the two modes {\it conditioned\/} 
on a single-photon count in the other.\hfill\break
{\bf Random state generation}: where the required state is `polluted'
by contributions from the vacuum state. In this case, the state may
be inferred by destructive photodetection, but then it cannot
leave the apparatus. For example, a weakly coupled four-mode 
down-converter~(\ref{4mode}) can make polarization entangled states 
{\it randomly\/} (in the sense given above); interestingly, it is 
currently unknown whether such states can be produced conditionally from 
the coherent states, linear optics, down-converters and photodetectors.

We see from these examples that the `cheap' non-linearity introduced by 
particle detection can increase the versatility of linear Bogoliubov
transformations.  However, there still appear to be limitations:

\noindent
{\bf Theorem (no-go for macro-superpositions)}: Detection of a single 
photon in one mode and no photons in any number of other modes cannot 
{\it conditionally\/} create superpositions of macroscopically distinct 
states given an initial vacuum state and using an arbitrarily complicated 
combination of linear multi-port interferometers, down-converters, 
squeezers, etc (all described by quadratic interactions).

\noindent
{\bf Proof}: Consider such a combination of components acting on the
vacuum. By Bloch-Messiah reduction (see Fig.~\ref{fig1}) the initial linear
multi-port interferometer described by $V^\dagger$ preserves the vacuum 
state, so only the later components have an affect. Since the
individual single-mode squeezers have evolution operators
which may be trivially normally ordered we may immediately write out the
general form for the outgoing Gaussian state as
\begin{equation}
|\psi_{\rm out}\rangle \propto \exp(\case{1}{2}\sum_{jk} B_{jk}
\hat b_j^\dagger \hat b_k^\dagger )\,|0\rangle \;,
\end{equation}
where without loss of generality $B_{jk}$ may be chosen to be complex
symmetric and $\hat b_j^\dagger$ are the outgoing mode creation operators.
Suppose a single photon is detected in some mode $\hat b_\ell$ and vacuum 
in several others, the conditioned state is
\begin{equation}
|\psi_{\rm cond}\rangle\propto {}_{\rm det}\langle 0|
\,\hat b_\ell\, |\psi_{\rm out}\rangle 
\propto {\sum_m}' B_{\ell m}\,\hat b_m^\dagger\; {}_{\rm det}\langle 0|
\psi_{\rm out}\rangle \label{cond} \;,
\end{equation}
where $|0\rangle_{\rm det}$ is the vacuum state for the subset of 
detected modes and the sum runs only over non-detected modes. It
is easy to see that ${}_{\rm det}\langle 0| \psi_{\rm out}\rangle$ is
a Gaussian state on the remaining modes, so the conditionally created
state from {\it single\/} photon detection is seen to be a sum of 
branches which differ by the placement of only a single photon in one 
mode or another.                                    \hfill $\Box$

\vskip 0.15truein

\noindent
{\bf Remark}:
Large amplitude coherent states are `macroscopic superpositions'
only in the sense that they are superpositions of macroscopic states
(although these states are not macroscopically distinguishable). Thus,
we have given a no-go theorem against creating so-called Schr\"odinger 
cat states for any such scheme without regard to the specific details of 
any particular implementation. A consequence of this result is that 
{\it entanglement\/} may not be `amplified' by say injecting microscopic 
superpositions into strongly pumped down-converters as has been recently
suggested by De Martini \cite{demart1} (though obviously superposition states 
may be sent through an amplifier \cite{damp}). A more detailed analysis 
of the specific scheme of Ref.~\onlinecite{demart1} supports the more 
general result of Eq.~(\ref{cond}) in our no-go theorem \cite{demart2}.

The Bloch-Messiah reduction theorem teaches us some important lessons about
the interconvertibility of different kinds of sources. For example, 
we find that a single squeezed state is an irreducible resource
which cannot be made from any number of lesser squeezed states and
linear optics. Similarly, if some device requires some given number
of squeezers in Bloch-Messiah reduced form then fewer squeezers plus linear 
optics will never suffice for the device's construction. Let us use
these observations to relate the three types of down-converters
S, D${}_2$ and E${}_4$.

\begin{figure}[thb]
\begin{psfrags}
\epsfxsize=3.2in
\psfrag{S}[bc]{~~S}
\psfrag{D2}[bc]{~\,D${}_2$}
\psfrag{BS}[c]{~~~BS}
\epsfxsize=3.2in
\epsfbox[-40 -20 450 125]{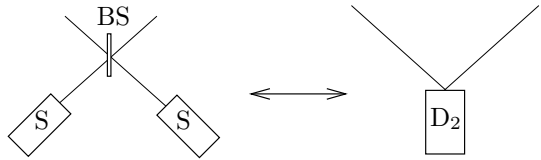}
\end{psfrags}
\caption{Bloch-Messiah equivalence: Here we illustrate the equivalence 
between pair of squeezers (S) combined at a 50:50 beam-splitter (BS) and 
a single two-mode down-converter (D${}_2$).}
\label{fig2}
\end{figure}

A non-entangling two-mode down-converter (D${}_2$) with 
coupling~(\ref{2mode}) requires two squeezers in reduced form as
is illustrated in Fig.~\ref{fig2}. For weak coupling this device is
a source of {\it random\/} photon pairs generated into distinct modes.
The Bloch-Messiah reduction into two squeezers and a 50:50 beam-splitter gives 
us a more sophisticated understanding of the Hong-Ou-Mandel interferometer 
\cite{HOM}. Away from the weak coupling limit we retrieve the twin-beam 
scheme for making two-mode squeezed states from a pair of independently 
squeezed states \cite{TB1}. Bloch-Messiah reduction neatly formalizes these 
multi-photon interference phenomena.

\begin{figure}[thb]
\begin{psfrags}
\psfrag{D2}[bc]{~~D${}_2$}
\psfrag{D4}[bc]{~~E${}_4$}
\psfrag{O}[bc]{$\odot$}
\psfrag{/}[c]{$\nearrow\mkern-17.6mu\swarrow$}
\psfrag{PBS}[c]{PBS~}
\epsfxsize=3.2in
\epsfbox[-30 -20 560 236]{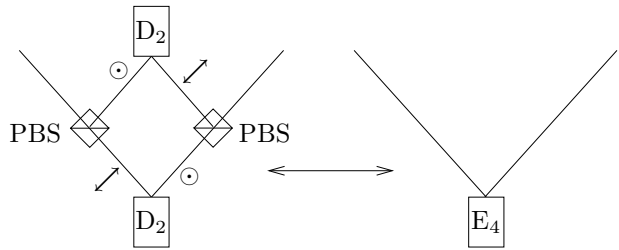}
\end{psfrags}
\caption{Polarization entanglement without loss of which-way information. 
Here we illustrate the equivalence between an entangling four-mode 
down-converter (E${}_4$) with a pair of non-entangling two-mode 
down-converters (D${}_2$) which are {\it randomly\/} creating photon pairs 
with opposite polarizations $\odot$, $\nearrow\mkern-17.6mu\swarrow$.
The polarization dependent beam-splitters (PBS) direct all photons
to the upper paths. Bloch-Messiah reduction shows that is impossible to 
(randomly) create such entangled states with only a {\it single\/} pass 
through a single non-entangling two-mode down-converter.}
\label{fig3}
\end{figure}

Similarly, Bloch-Messiah reduction applied to the entangling four-mode
down-converter (E${}_4$) of Eq.~(\ref{4mode}) shows that four squeezers 
are required in reduced form. Thus, a random polarization entangled state 
cannot be formed from a {\it single\/} pass through a single 
non-entangling down-converter [D${}_2$, Eq.~(\ref{2mode})]. Nonetheless, 
it may be made easily enough with two such devices \cite{111}. In 
Fig.~\ref{fig3} we give just such an equivalence. Curiously, this 
construction produces entanglement without erasing the which-way information 
about the photons. It should be noted that this scheme is very different 
(in terms of the irreducible resources used) than the entanglement swapping 
scheme of Zukowski et al \cite{ES} which starts with a pair of entangling 
down-converters.

As a final application for the Bloch-Messiah reduction theorem we consider
constructing optimal optical circuits using as little squeezing as
possible. Consider the ideal quantum non-demolition (QND) coupling 
between a pair of optical quadrature-phase amplitudes
\begin{eqnarray}
\hat b_1 &=& \hat a_1 -\case{1}{2}\hat a_2 +\case{1}{2}\hat a_2^\dagger 
\nonumber \\
\hat b_2 &=& \case{1}{2}\hat a_1 +\hat a_2 +\case{1}{2}\hat a_1^\dagger 
\label{QND} \;.
\end{eqnarray}
The relevant decomposition is given by
\begin{eqnarray}
A&=&\!\left(\begin{array}{rr}\sin\theta &-i\cos\theta\\
\cos\theta& i\sin\theta\end{array}\right)\!
\left(\begin{array}{cc}\case{\sqrt{5}}{2}&0\\
0&\case{\sqrt{5}}{2}\end{array}\right) \!
\left(\begin{array}{rr}\cos\theta &-i\sin\theta\\
\sin\theta& i\cos\theta\end{array}\right)^\dagger \nonumber \\
B&=&\!\left(\begin{array}{rr}\sin\theta &-i\cos\theta\\
\cos\theta& i\sin\theta\end{array}\right)\!
\left(\begin{array}{cc}\case{1}{2}&0\\
0&\case{1}{2}\end{array}\right) \!
\left(\begin{array}{cc}\cos\theta &-i\sin\theta\\
\sin\theta& i\cos\theta\end{array}\right)^{\rm T} \!\!\!\!, \!\label{degen}
\end{eqnarray}
where $\theta=\case{1}{2}\sin^{-1}(2/\sqrt{5})\simeq 31.72^\circ$. 
The circuit consists of a pair of squeezers with equal squeezing parameters
of $r = \ln[(1+\sqrt{5})/2]$ (corresponding to roughly $4.18$~dB) and
a pair of unequal unbalanced beam-splitters with energy transmission
coefficients of $27.64\%$ and $72.36\%$. 

In fact, this circuit is equivalent to one derived by Yurke \cite{Yurke}, 
however, Bloch-Messiah reduction guarantees its optimality. We can improve 
on it further by noting that the singular value eigenvalues in~(\ref{degen})
are degenerate and so the decomposition is not unique; a construction
with much simpler 50:50 beam-splitters is given by
$$ 
U=\case{1}{\sqrt{2}}\left(\begin{array}{rr}
i e^{i\theta} & i e^{-i\theta} \\
- e^{i\theta}&   e^{-i\theta} \end{array}\right) ~~~~
V=\case{1}{\sqrt{2}}\left(\begin{array}{rr}
- e^{-i\theta} &   e^{i\theta} \\
-i e^{-i\theta}& -i e^{i\theta} \end{array}\right) \;,
$$ 
with $\theta$ as above.  We note that the QND coupling~(\ref{QND}) 
has recently been used in error correction codes for quantum optical 
fields \cite{Braunstein,Lloyd}. 

In conclusion, we have illustrated the utility of the Bloch-Messiah reduction 
theorem for linear bosonic Bogoliubov transformations in the context
of quantum optics. We have shown the equivalence between a number of 
elementary sources of weak random states, including a simple scheme to 
randomly generate polarization entanglement without a loss of which-way 
information. When supplemented by detection of a single photon we have 
shown that superpositions of macroscopically distinct states cannot be 
created out of vacuum using linear optics and down-converters, squeezers, 
etc (all corresponding to linear Bogoliubov transformations). Finally, we 
used Bloch-Messiah reduction to study the construction of minimal optical 
circuits. Although we have concentrated on applications for photonic modes 
in quantum optics the Bloch-Messiah reduction theorem holds for all bosonic 
modes.

\vskip 0.15truein

\noindent
{\bf Appendix: Proof of Bloch-Messiah reduction}
\vskip 0.1truein

Without loss of generality, we may 
set the displacements in Eq.~(\ref{BT}) to zero, i.e., $\beta_j=0$. The 
canonical commutation relations for $\hat b_j$ in Eq.~(\ref{BT}) impose 
the conditions \cite{Ekert}
\begin{eqnarray}
A B^{\rm T} &=& (A B^{\rm T})^{\rm T} \label{rel1} \\
A A^\dagger&=&B B^\dagger +{\openone} \label{rel2} \;,
\end{eqnarray}
since $A A^\dagger$ and $B B^\dagger$ are hermitian and
according to Eq.~(\ref{rel2}) must commute, they also may be diagonalized
in the same basis by some unitary matrix $U$. However, using the singular
value decomposition theorem \cite{SVDthm} we can always diagonalize
$A=UA_DV^\dagger$ and $B=UB_DW^\dagger$ into non-negative matrices $A_D$
and $B_D$ satisfying Eq.~(\ref{1moderel}) where $V$ and $W$ are a
pair of unitary matrices. Unitarity of~(\ref{BT}) guarantees a unique
inverse which with the aid of Eqs.~(\ref{rel1}) and~(\ref{rel2}) may be
easily computed to be \cite{Ekert}
\begin{equation}
\hat a_j = \sum_k (A_{kj}^\ast \hat b_k -B_{kj} \hat b^\dagger_k)
\label{invBT} \;.
\end{equation}
Imposing the canonical commutation relations again here yields
the conditions
\begin{eqnarray}
A^\dagger B &=& (A^\dagger B)^{\rm T} \label{rel3} \\
A^\dagger A&=&(B^\dagger B)^{\rm T} +{\openone} \label{rel4} \;.
\end{eqnarray}
Thus we see that $A^\dagger A$ and $(B^\dagger B)^{\rm T}$ may be
diagonalized in the same basis by a unitary matrix $V=W^\ast$ which
yields Eq.~(\ref{glub}) as required. Finally, we note that this form
for $A$ and $B$ automatically satisfies the subsidiary conditions of
Eqs.~(\ref{rel1}) and (\ref{rel3}).                     \hfill$\Box$




\end{document}